\documentclass[10pt]{article}
\usepackage{url}
\usepackage{latexsym}
\usepackage{hyperref}
\usepackage{graphicx}
\usepackage{multirow}
\usepackage[usenames,dvipsnames]{color}

\usepackage{cite}

\usepackage{pdflscape}

\newcommand{\benum}{\begin{enumerate}}
\newcommand{\eenum}{\end{enumerate}}
\newcommand{\bitem}{\begin{itemize}}
\newcommand{\eitem}{\end{itemize}}
\newcommand{\bCitem}
	   {\begin{list}
	    {$\bullet$}
	    {\itemsep 0.01in \parsep -.01in \topsep .08in}}
\newcommand{\eCitem}{\end{list}}

\newlength{\oldbaselineskip}

\newcommand{\comment}[1]{}

\newcommand{\dictentry}[1]{\vspace{1 cm}}

\def\scaledpicture #1 by #2 (#3) indent #4 scale #5 {}

\title{A Comparison of the Pathway Tools Software with the \\Reactome Software}

\author{Peter D. Karp\\
Bioinformatics Research Group\\
SRI International, Menlo Park, USA\\
pkarp@ai.sri.com\\
}

\begin{document}

\maketitle

\section*{Abstract}

This document compares SRI's Pathway Tools (PTools) software with the Reactome software.  Both software systems serve the pathway bioinformatics area, including representation and analysis of metabolic pathways and signaling pathways.  The comparison covers pathway bioinformatics capabilities, but does not cover other major facets of Pathway Tools that are completely absent from the Reactome software: Pathway Tools genome-informatics capabilities, regulatory informatics capabilities, and table-based analysis tools (SmartTables).

Our overall findings are as follows.  (1) PTools is significantly ahead of Reactome in its basic information pages. For example, PTools pathway layout algorithms have been developed to an advanced state over several decades, whereas Reactome pathway layouts are illegible, omit important information, and are created manually and therefore cannot scale to thousands of genomes.  (2) PTools is far ahead of Reactome in omics analysis.  PTools includes all of the omics-analysis methods that Reactome provides, and includes multiple methods that Reactome lacks.  (3) PTools contains a metabolic route search tool (searching for paths through the metabolic network), which Reactome lacks.  (4) PTools is significantly ahead of Reactome in inference of metabolic pathways from genome information to create new metabolic databases.  (5) PTools has an extensive complement of metabolic-modeling tools whereas Reactome has none.  (6) PTools is more scalable than Reactome, handling 18,000 genomes versus 90 genomes for Reactome.  (7) PTools has a larger user base than Reactome. PTools powers 17 websites versus two for Reactome.  PTools has been licensed by 10,800 users (Reactome licensed user count is unknown).

\section{Introduction}

This document compares SRI's Pathway Tools software
\cite{PTools19BiB,PTools19Overview} with the Reactome software
\cite{Reactome20}.  Both software systems serve the pathway
bioinformatics area, including representation and analysis of
metabolic pathways and signaling pathways.  The comparison covers
pathway bioinformatics capabilities, but does not cover other major
facets of Pathway Tools that are completely absent from the Reactome
software: Pathway Tools genome-informatics capabilities, regulatory
informatics capabilities, and table-based analysis tools
(SmartTables).

Pathway Tools (PTools) powers the BioCyc.org website and 16 other pathway
websites \cite{MetaCycOtherPathwayDBsURL}.  The Reactome
software powers Reactome.org and plantreactome.gramene.org.

This comparison is divided into three sections:

\bitem
\item Comparison of pathway pages, reaction pages, and metabolite pages

\item Comparison of pathway-informatics analysis tools

\item Comparison of metabolic modeling tools
\eitem

Pathway Tools has been shown to scale to 18,000 genomes (biocyc.org).  Reactome has been shown to scale to 90 genomes
(reactome.org).

\section{Pathway, Reaction, and Metabolite Pages}

\begin{figure}
  \begin{center}
    \includegraphics[width=5.5in]{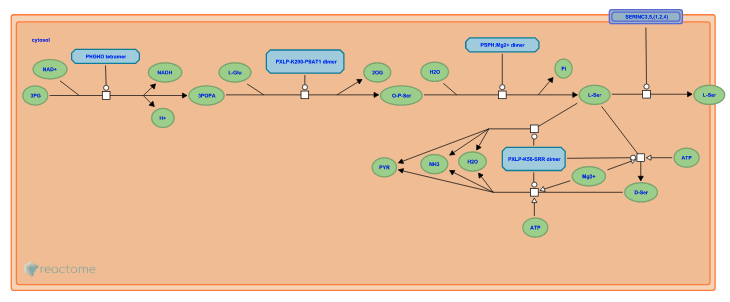}
    \caption{Reactome pathway page for serine biosynthesis pathway.
      The legibility of this figure is quite similar when this PDF is
      viewed at 100\% magnification to how the pathway diagram appears in the Reactome web page.}
\label{fig:serine1-reactome}
  \end{center}
\end{figure}

The most basic and essential aspect of a pathway database and website
is the ability to  legibly depict meaningful pathway diagrams.  Reactome
fails in this regard: pathway diagrams shown in Reactome pathway
pages are not legible (i.e., metabolite names are not readable in
these diagrams), even for 
a small pathway such as serine biosynthesis (see Figure~\ref{fig:serine1-reactome}
(\url{https://reactome.org/content/detail/R-CFA-977347}).  Larger
pathways are even harder to make out.
If the user clicks on the pathway diagram in the preceding Reactome page to display the same diagram within
the ``pathway browser'', the diagram can then be zoomed to a 
magnification where the metabolite names are legible.  However, the
metabolite names shown are abbreviations, many of which will not be
meaningful to most users.  (Most likely the abbreviations are being
used in an effort to make the pathway diagrams smaller.)

\begin{figure}
  \begin{center}
    \includegraphics[width=7.5in]{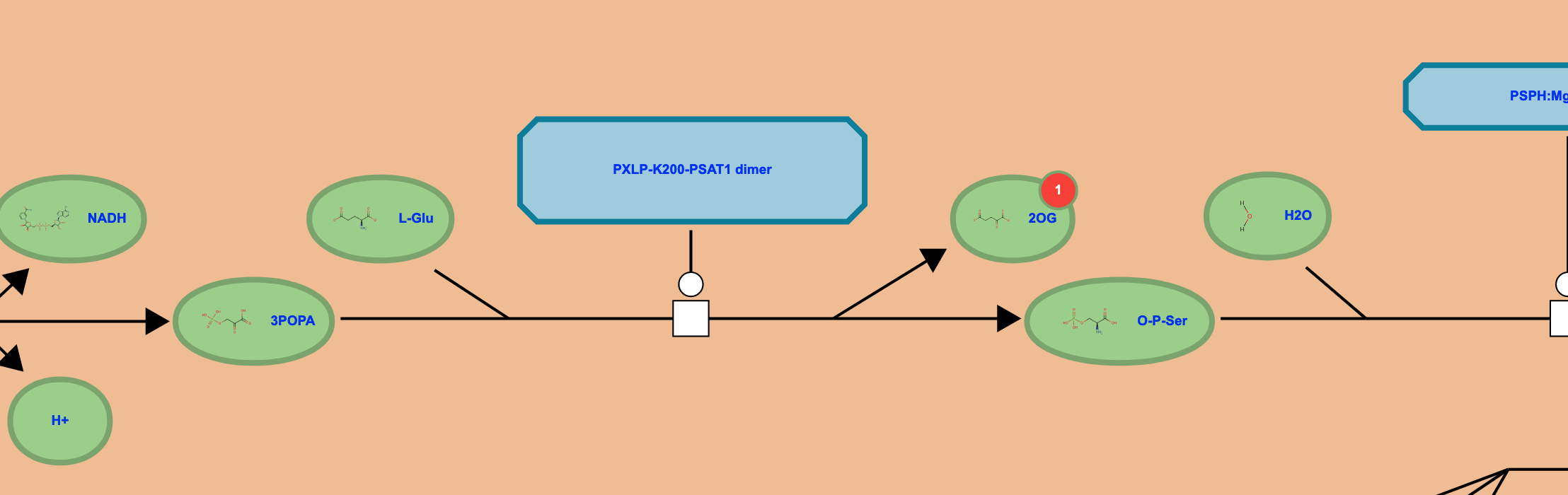}
    \caption{Reactome pathway browser, serine biosynthesis pathway,
      zoomed until chemical structures are visible. The legibility of
      this figure is quite similar to that of the Reactome web page.}
\label{fig:serine2-reactome}
  \end{center}
\end{figure}

Further, Reactome pathway diagrams lack enzyme
names, gene names, and EC numbers --- all of which are important
facets of communicating the pathway.  Magnifying this diagram eventually
results in depiction of metabolite chemical structures, although at
this point the diagram must be zoomed so much that only one reaction
from the pathway is visible, and the chemical structures are barely
legible (see Figure~\ref{fig:serine2-reactome}).

\begin{figure}
  \begin{center}
    \includegraphics[width=6.5in]{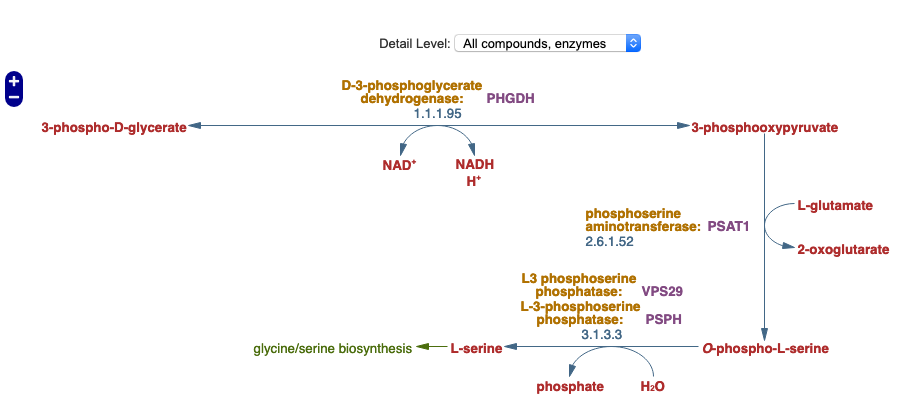}
    \caption{BioCyc serine biosynthesis pathway page generated by
      Pathway Tools; the legibility of this figure is quite similar
      when this PDF is viewed at 100\% magnification to how the
      pathway diagram appears in the BioCyc web page.}
\label{fig:serine1-ptools}
  \end{center}
\end{figure}

\begin{figure}
  \begin{center}
    \includegraphics[width=6.5in]{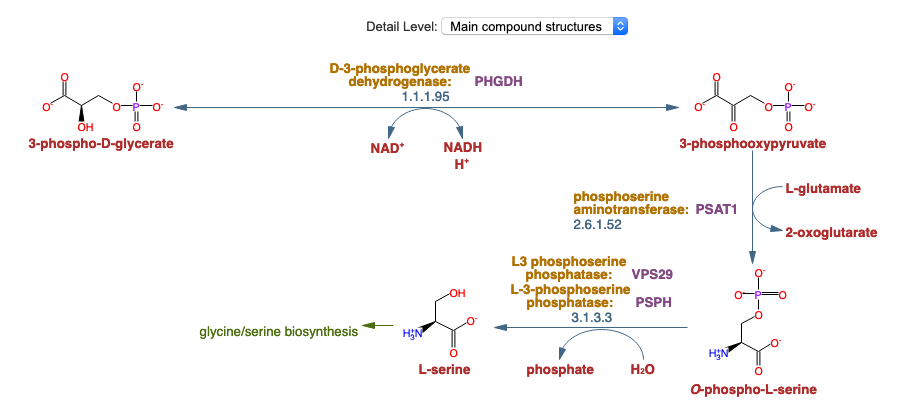}
    \caption{BioCyc serine biosynthesis pathway page, showing chemical structures.}
\label{fig:serine2-ptools}
  \end{center}
\end{figure}

In contrast, the HumanCyc L-serine biosynthesis pathway produced by
Pathway Tools
(\url{https://biocyc.org/HUMAN/NEW-IMAGE?type=PATHWAY&object=SERSYN-PWY})
is easily legible both with and without chemical structures (see
Figures~\ref{fig:serine1-ptools} and \ref{fig:serine2-ptools}), and
the PTools diagram can
be customized to depict any combination of gene names, enzyme names,
EC numbers, and chemical structures.

Comparisons between Reactome and HumanCyc signaling pathways are
similar in terms of legibility.  The Reactome pathway for BMP
Signaling is illegible (\url{https://reactome.org/content/detail/R-HSA-201451}),
whereas the HumanCyc pathway for BMP Signaling is easily legible
(\url{https://humancyc.org/HUMAN/NEW-IMAGE?type=PATHWAY&object=PWY66-11}).

\begin{figure}
  \begin{center}
    \includegraphics[width=4.5in]{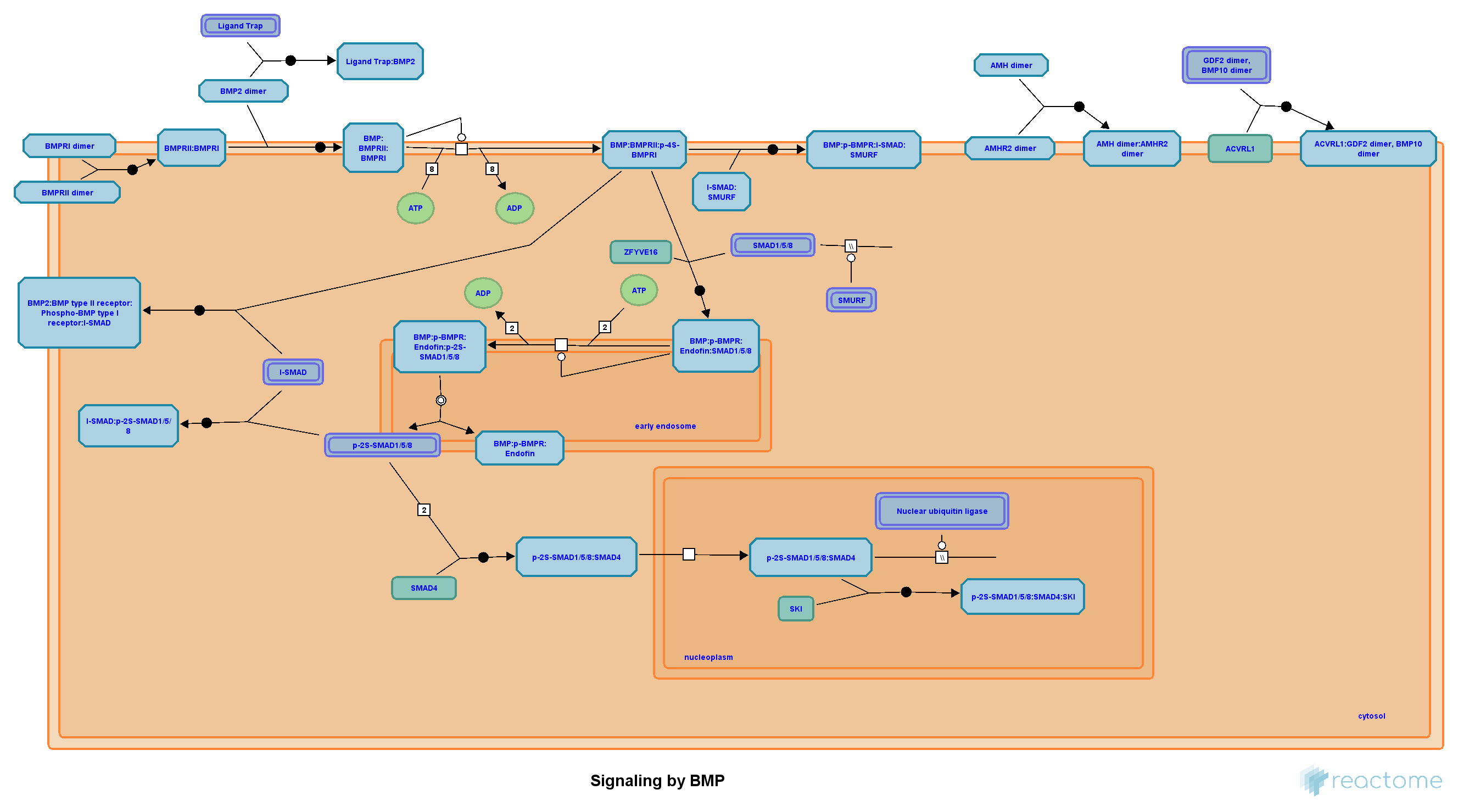}
    \caption{Reactome BMP Signaling pathway.  The legibility of this figure is quite similar
      when this PDF is viewed at 100\% magnification to how the
      pathway diagram appears in the Reactome web page.}
\label{fig:bmp-reactome}
  \end{center}
\end{figure}
\begin{figure}
  \begin{center}
    \includegraphics[width=7.5in]{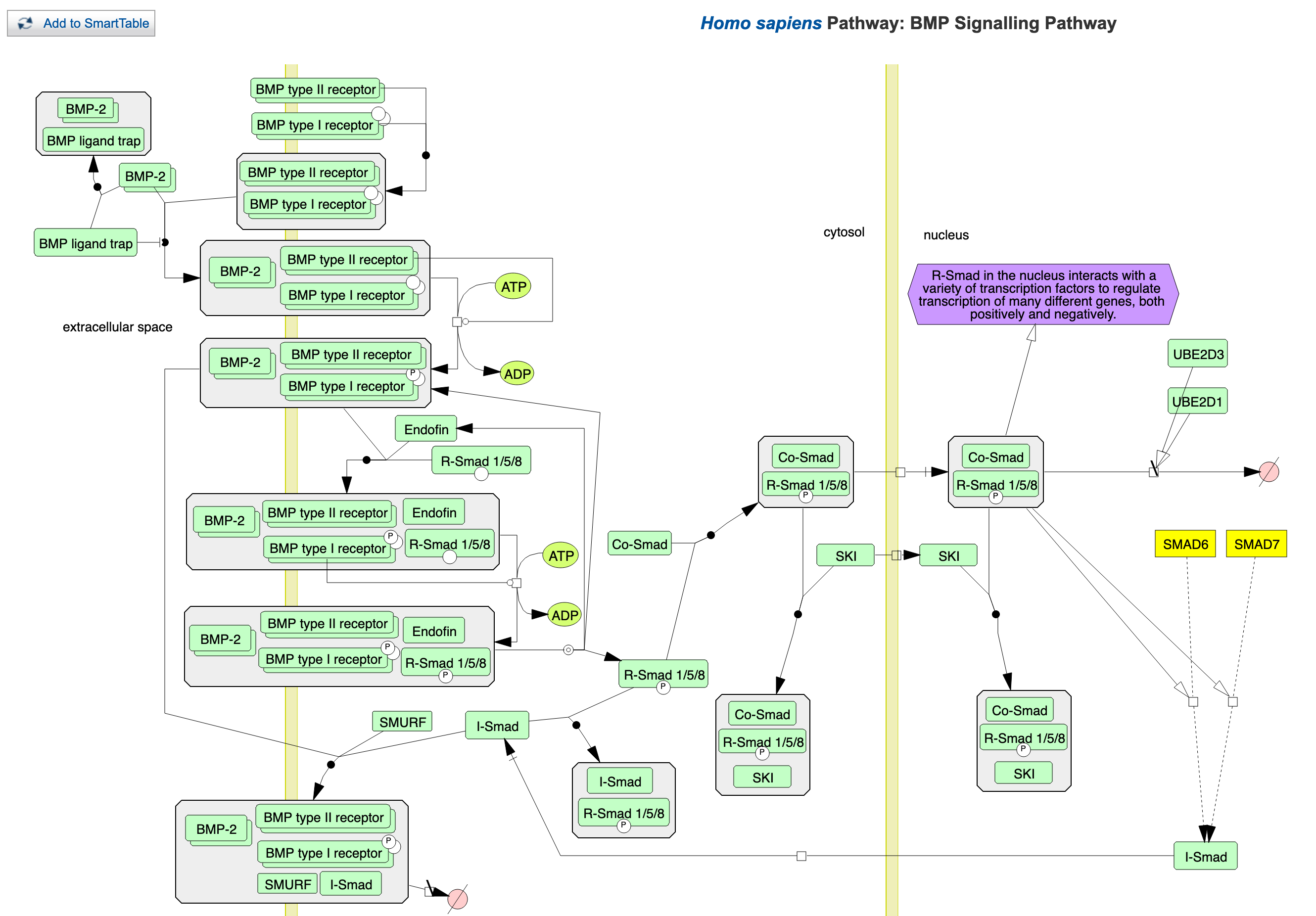}
    \caption{BioCyc BMP Signaling pathway generated by Pathway Tools.
The legibility of this figure is quite similar
      when this PDF is viewed at 100\% magnification to how the
      pathway diagram appears in the BioCyc web page.
}
\label{fig:bmp-ptools}
  \end{center}
\end{figure}

Reactome provides two full-metabolic-network diagrams.  The ``pathway browser''
(\url{https://reactome.org/PathwayBrowser/}) shows a set of cascading
circular diagrams for different pathway categories.  However, the
diagram essentially conveys no information other than the grouping of pathways
in functional categories.  In contrast, the PTools Cellular
Overview diagram
(\url{https://humancyc.org/overviewsWeb/celOv.shtml?orgid=HUMAN})
not only groups related pathways together, it also
depicts the reactions and metabolites within a pathway at high zoom levels; the diagram
also supports painting of omics data onto the pathway diagrams, with
animations used to depict multiple time points.  This PTools diagram
can also be searched by pathway, gene, enzyme, and metabolite names.
The second Reactome network diagram, the ``Voronoi diagram'' \cite{Reactome20},
is simply a hierarchical listing of all Reactome pathways without the pathway
diagrams; it conveys essentially
no information, is not searchable, and does not support navigation to
individual pathway diagrams.

\begin{table}[!h]
\centerline{
\begin{tabular}{|l|c|c|c|c|c|c|c|} \hline
{\bf Tool} & {\bf PTools}                 & {\bf Reactome} \\ \hline \hline
Metabolic Pathway Page                & YES        & YES \\ \hline
-- Depict Full Metabolite Names       & YES        & no  \\ \hline
-- Depict Metabolite Structures       & YES        & illegible \\ \hline
-- Depict Enzyme Names                & YES        & no  \\ \hline
-- Depict EC numbers                  & YES        & no  \\ \hline
-- Depict Compartment Information     & YES        & YES  \\ \hline
-- Depict Enzyme Regulation           & YES        & YES \\ \hline
-- Depict Genetic Regulation          & YES        & YES \\ \hline
-- Automatic Pathway Layout           & YES        & no  \\ \hline
-- Paint Omics Data onto Pathway      & YES        & YES \\ \hline
Customizable Multi-Pathway Diagram    & YES        & no  \\ \hline
-- Paint Omics Data onto Multi-Pathway & YES       & no \\ \hline
Signaling Pathway Page                & YES        & YES \\ \hline
Metabolite Page                       & YES        & no \\ \hline
Reaction Page                         & YES        & no \\ \hline
-- Reaction Atom Mappings             & YES        & no \\ \hline
Interactive Metabolic Pathway Editor  & YES        & YES \\ \hline
Interactive Signaling Pathway Editor  & YES        & YES \\ \hline
Interactive Reaction Editor           & YES        & YES \\ \hline
Interactive Metabolite Editor         & YES        & no \\ \hline
\end{tabular}
}
\caption{\label{tab:pathway-page}
{\bf Pathway Page Comparison.}
}
\end{table}

Table~\ref{tab:pathway-page} assesses the capabilities of
information pages for pathways, reactions, and metabolites in PTools
and Reactome, as well as interactive editors associated with pathways,
reactions, and metabolites.

An explanation of the rows within Table~\ref{tab:pathway-page} is as follows.

\begin{itemize}

\item {\bf Pathway Page}: Is a metabolic-pathway information page with pathway
  diagram provided?

\item {\bf Depict Full Metabolite Names}: Do pathway diagrams
  include meaningful metabolite names (as opposed to abbreviations
  such as ``O-P-Ser'' or ``3POPA'')?

\item {\bf Depict Metabolite Structures}: Can pathway diagrams
  show the chemical structures of metabolites?

\item {\bf Depict Enzyme Names}: Do pathway diagrams
  include enzyme names?

\item {\bf Depict EC Numbers}: Do pathway diagrams
  include EC numbers?

\item {\bf Depict Compartment Information}: Do pathway diagrams
  depict cellular compartments and membranes?

\item {\bf Depict Enzyme Regulation}: Can pathway diagrams
   show regulation of enzymes by metabolites, to depict information such as feedback inhibition?

\item {\bf Depict Genetic Regulation}: Can pathway diagrams
   show genetic regulation of enzymes, such as by transcription
   factors and attenuation?

\item {\bf Automatic Pathway Layout}: Are pathway diagrams generated
  automatically by the software, thereby avoiding manual drawing, and
  enabling scalability to thousands of organisms?

\item {\bf Paint Omics Data onto Pathway}: Can a user visualize 
   omics data on pathway diagrams?

\item {\bf Customizable Multi-Pathway Diagram}: Can users interactively create
  diagrams consisting of multiple interacting metabolic pathways?

\item {\bf Paint Omics Data onto Multi-Pathway}: Can a user visualize 
   omics data on multi-pathway diagrams?

\item {\bf Signaling Pathway Page}: Is a signaling-pathway information page with pathway
  diagram provided?

\item {\bf Metabolite Page}: Does the site provide a metabolite page,
  showing relevant information such as synonyms, chemical structure,
  and reactions in which the metabolite occurs?

\item {\bf Reaction Page}: Does the site provide a reaction page,
  showing relevant information such as EC numbers, reaction equation,
  and enzymes catalyzing the reaction?

\item {\bf Reaction Atom Mappings}: Can the reaction equation
  be shown with metabolite structures that depict the trajectories of atoms
  from reactants to products?

\item {\bf Interactive Metabolic Pathway Editor}: Does the software provide an
  editor for interactive creation and modification of metabolic pathways?

\item {\bf Interactive Signaling Pathway Editor}: Does the software provide an
  editor for interactive creation and modification of signaling pathways?

\item {\bf Interactive Reaction Editor}: Does the software provide an
  editor for interactive creation and modification of reactions?

  \item {\bf Interactive Metabolite Editor}: Does the software provide an
  editor for interactive creation and modification of metabolites?

\end{itemize}

\section{Pathway Informatics Tools}

Table~\ref{tab:metabolic-tools} assesses a number of analysis
capabilities within the two software packages.

\begin{table}[!h]
\centerline{
\begin{tabular}{|l|c|c|} \hline
{\bf Tool} & {\bf PTools}                 & {\bf Reactome} \\ \hline \hline
Full Metabolic Network Diagram        & YES        & YES \\ \hline
-- Zoomable Metabolic Network         & YES        & YES \\ \hline
-- Paint Omics Data onto Diagram      & YES        & no  \\ \hline
-- Animated Omics Data Painting       & YES        & no  \\ \hline
-- Metabolic Poster                   & YES        & no  \\ \hline
Metabolic Reconstruction              & YES        & YES \\ \hline
Route Search Tool                     & YES        & no \\ \hline
Genome-Scale Reactome Comparison      & YES        & no \\ \hline
Genome-Scale Pathway Comparison       & YES        & no \\ \hline
Transcriptomics Enrichment Analysis   & YES        & YES \\ \hline
Metabolomics Enrichment Analysis   & YES        & YES \\ \hline
Metabolomics Pathway Covering Analysis   & YES        & YES \\ \hline
\end{tabular}
}
\caption{\label{tab:metabolic-tools}
{\bf Pathway Informatics Tools Comparison.}
}
\end{table}

\begin{itemize}
\item {\bf Full Metabolic Network Diagram}: Can the entire metabolic
  reaction network of an organism be depicted and explored by an interactive graphical
  interface?

\item {\bf Zoomable Metabolic Network}: Does the metabolic network
  browser enable real-time semantic zooming of the network?

\item {\bf Paint Omics Data onto Network}: Can a user visualize 
  an omics dataset (e.g., gene expression, metabolomics) on the
  metabolic network diagram?

\item {\bf Animated Omics Data Painting}: Can several omics measurements be
  visualized as an animation on the metabolic network diagram?

\item {\bf Metabolic Poster}: Can the portal generate a printable
  wall-sized poster of the organism's metabolic network?

\item {\bf Metabolic Reconstruction}: Starting from a
  functionally annotated genome, can the software infer the organism's
  metabolic reaction network and pathways?  Reactome apparently has an
  algorithm for this, but it has never been published.  Further,
  because Reactome does not have a reference pathway database that
  spans all domains of life as does the MetaCyc \cite{MetaCycNAR20} database used by
  PTools for metabolic reconstruction, Reactome cannot accurately
  predict pathways across all domains of life as can PTools.

\item {\bf Route Search Tool}: Given a starting and an ending
  metabolite, can the site compute an optimal series
  of known reactions (routes) that converts the starting metabolite to the
  ending metabolite?
  
\item {\bf Transcriptomics Enrichment Analysis}: Can the site compute 
   statistical enrichment of pathways from transcriptomics data?

\item {\bf Metabolomics Enrichment Analysis}: Can the site compute 
   statistical enrichment of pathways from metabolomics data?

\item {\bf Metabolomics Pathway Covering Analysis}: Can the site compute 
   a minimal set of metabolic pathways that cover a set of metabolites
   from a metabolomics experiment?

\end{itemize}

\section{Metabolic Modeling}

Table~\ref{tab:modeling} assesses metabolic-modeling
capabilities within the two software packages.

\begin{table}[!h]
\centerline{
\begin{tabular}{|l|c|c|} \hline
{\bf Tool} & {\bf PTools}                 & {\bf Reactome} \\ \hline \hline
Execute Metabolic Model               & YES        & no  \\ \hline
-- Gene Knock-Out Analysis            & YES        & no  \\ \hline
-- Flux-Variability Analysis          & YES        & no  \\ \hline
Model Organism Communities            & YES        & no  \\ \hline
Reaction Gap Filling                  & YES        & no  \\ \hline
Chokepoint Analysis                   & YES        & no  \\ \hline
Dead-End Metabolite Analysis          & YES        & no  \\ \hline
Blocked-Reaction Analysis             & YES        & no  \\ \hline
\end{tabular}
}
\caption{\label{tab:modeling}
{\bf Metabolic Modeling Comparison.}
}
\end{table}

\begin{itemize}

\item {\bf Execute Metabolic Model}: Can a user execute a steady-state
  metabolic flux model using the flux-balance analysis approach?

\item {\bf Gene Knock-Out Analysis}: Can a user run flux-balance analysis (FBA) on the metabolic
  network by systematically disabling (knocking-out) various genes, to investigate
  how knock-outs perturb the network, and to predict gene essentiality?

\item {\bf Flux-Variability Analysis}: Can a user run flux-variability
  analysis to compute the range of fluxes each reaction can attain?

\item {\bf Model Organism Communities}: Does the software enable
  modeling of organism communities as well as single organisms?

\item {\bf Reaction Gap Filling}: Does the software have a tool for
  automatically proposing reactions to add to the model to fill gaps
  in the metabolic network?

\item {\bf Chokepoint Analysis}: Can the site compute chokepoint
  reactions (possible drug targets) in the full metabolic reaction network?  A chokepoint
  reaction is a reaction that either uniquely consumes a specific
  reactant or uniquely produces a specific product in the metabolic
  network. 

\item {\bf Dead-End Metabolite Analysis}: Can the portal compute dead-end
  metabolites in the full metabolic reaction network?  Dead-end
  metabolites are those that are either only consumed, or only
  produced, by the reactions within a given cellular compartment,
  including transport reactions.

\item {\bf Blocked-Reaction Analysis}: Can the portal compute blocked
  reactions in the full metabolic reaction network?  Blocked reactions
  cannot carry flux because of dead-end metabolites upstream or
  downstream of the reactions.

%%%%\item {\bf Metabolite Tracing Tool}: Can a user navigate the metabolic
%%%%  reaction network in an interactive fashion, moving from a starting
%%%%  metabolite along selectable reactions to a series of further
%%%%  metabolites?

\end{itemize}

\section{Conclusions}

Our overall findings are as follows.

\begin{itemize}
\item PTools is significantly ahead of Reactome in its
basic information pages. For example, PTools pathway layout algorithms
have been developed to an advanced state over several decades, whereas
Reactome pathway layouts are illegible, omit important information,
and are created manually and therefore cannot
scale to thousands of genomes.

\item PTools includes a full metabolic network diagram that includes
  real-time semantic zooming and is far ahead of the comparable tools
  in Reactome.

\item PTools is far ahead of Reactome in omics analysis.  PTools includes all of the omics-analysis methods that
Reactome provides, and includes multiple methods that Reactome lacks,
such as an Omics Dashboard and depiction of omics data on a zoomable metabolic
map diagram.

\item PTools contains a
metabolic route search tool (searching for paths through the metabolic
network), which Reactome lacks.

\item PTools is significantly ahead of Reactome in inference
  of metabolic pathways from genome information to create new
  metabolic databases.

\item PTools has an extensive complement of metabolic-modeling tools whereas Reactome has none.

\item PTools is more scalable than Reactome, handling 18,000
genomes versus 90 genomes for Reactome.

\item PTools has a larger user base than Reactome.  PTools powers 17
websites versus two for Reactome.  PTools has been licensed by
10,800 users (Reactome licensed user count is unknown).

\end{itemize}

If the Reactome data were made available via PTools, the accessibility
of those data would be greatly enhanced through legible pathway
diagrams.  Further, users could apply a wider variety of analysis
tools and operations to those data, such as for omics data analysis,
metabolic route searching, and metabolic modeling.

\bibliography{all}
\bibliographystyle{plain}

\end{document}